\documentclass[
 aip,
 jmp,%
 amsmath,amssymb,
 preprint,%
]{revtex4-1}

\usepackage{dcolumn}
\usepackage{bm}

\usepackage{amsfonts}    
\usepackage{amssymb}
\usepackage{amsmath}
\usepackage{latexsym}
\usepackage{eepic}
\usepackage{graphicx}
\usepackage[usenames,dvipsnames]{color}
\usepackage{color}

\newtheorem{conjecture}{Conjecture}

\begin{document}


\title[{\footnotesize}]{General $N^{th}$-order superintegrable systems separating in polar coordinates}

\author{A. M. Escobar-Ruiz}
\email{escobarr@crm.umontreal.ca}
\affiliation{
Centre de recherches math\'ematiques
and D\'epartement de math\'ematiques \\
et de statistique, Universit\'e de Montreal,  C.P. 6128,
succ. Centre-ville,\\ Montr\'eal (QC) H3C 3J7, Canada}

\author{P. Winternitz}
\email{wintern@crm.umontreal.ca}
\affiliation{
Centre de recherches math\'ematiques
and D\'epartement de math\'ematiques \\
et de statistique, Universit\'e de Montreal,  C.P. 6128,
succ. Centre-ville,\\ Montr\'eal (QC) H3C 3J7, Canada}

\author{\.{I}. Yurdu\c{s}en}
\email{yurdusen@hacettepe.edu.tr}
\affiliation{
Centre de recherches math\'ematiques
and D\'epartement de math\'ematiques \\
et de statistique, Universit\'e de Montreal,  C.P. 6128,
succ. Centre-ville,\\ Montr\'eal (QC) H3C 3J7, Canada}
\affiliation{Department of Mathematics, Hacettepe University,
                     06800 Beytepe,  \\ Ankara, Turkey}

\date{\today}

\begin{abstract}
The general description of superintegrable systems with one polynomial integral of order $N$ in the momenta is presented for a Hamiltonian system in two-dimensional Euclidean plane. We consider classical and quantum Hamiltonian systems allowing separation of variables in polar coordinates. The potentials can be classified into two major classes and their main properties are described. We conjecture that a new infinite family of superintegrable potentials in terms of the sixth Painlev\'e transcendent $P_6$ exists and demonstrate this for the first few cases.
\end{abstract}

\maketitle

\section{INTRODUCTION}
\label{intro}
Several recent articles were devoted to superintegrable systems in the
Euclidean space $E_2$ that allow the separation of variables in polar
\cite{TPW2010, AMJVPW2015, PostWinternitz:2010, TTWquantum, TTWclassical, AMJVPWIY2018} or Cartesian
\cite{gW, g, MSW, AW} coordinates and
admit an additional integral $Y$ of order $N$ with $2 \leq N \leq 5$.

In this article we concentrate on systems that are separable in polar coordinates
and posses an additional integral of arbitrary order $N$. The systems are
second-order integrable because in addition to the Hamiltonian
\begin{eqnarray}
{H}\  &=& \ -\frac{\hbar^2}{2}\,\bigg(\partial^2_r + \frac{1}{r}\partial_r + \frac{1}{r^2}\partial_\theta^2\bigg)
\ + \ V(r, \theta)\ ,
\qquad
V(r, \theta) =  R(r) \, + \, \frac{S(\theta)}{r^2}\ ,
\label{Hpolar}
\end{eqnarray}
they allow a second-order integral
\begin{eqnarray}
X\ &=& \ L_z^2 + 2 \,S(\theta) \ ,
\qquad \quad
x = r \cos \theta\,, \,\,\,y = r \sin \theta\,.
\label{X}
\end{eqnarray}

The existence of the integral $Y$, an $N^{th}$-order
polynomial in the momentum components, makes the system superintegrable
(more integrals of motion than degrees of freedom). In classical
mechanics $H$, $X$ and $Y$ are well-defined functions on phase
space and are functionally independent. In quantum mechanics
they are assumed to be polynomials, or convergent series in the
enveloping algebra of the Heisenberg algebra in $E_2$, {\it i.e.};
$\{x, y, p_x, p_y, 1\}$. The operators $H$, $X$, $Y$ are assumed to be
polynomially independent, {\it i.e.}; no Jordan polynomial in the quantities $(H, X, Y)$ is
equal to zero. We will use the usual vector fields spanning $e_2$
$(p_{x} = - i \hbar \partial_x,\,  p_{y} =- i \hbar \partial_y,\,  L_z=x\,p_y  -  y\,p_x)$
in quantum mechanics. In classical mechanics, the Hamiltonian is $H=\frac{1}{2}(p_r^2+\frac{1}{r^2}L_z^2)+ V(r,\,\theta)$, $p_r$ and $L_z$ are the components of linear momentum canonically conjugate to the coordinates $r$ and $\theta$, respectively.

For recent reviews of classical and quantum superintegrable systems see \cite{MillerPostWinternitz:2013, Millerebook}.

The aim of this article is to establish some general properties of
superintegrable systems separating in polar coordinates and allowing
a higher-order integral. The properties were observed for specific
choices of $N$. The results presented here will be generalizations to all
$N$ and presented ideally as theorems, otherwise as conjectures.

Among the results observed for $3\leq N \leq 5$ we mention
\begin{enumerate}
\item Superintegrable Hamiltonians in classical and quantum
mechanics can differ \cite{Hietarinta1998, HietarintaGrammaticos}. Terms depending on $\hbar$ can appear
in the quantum case. The classical limit $\hbar \rightarrow 0$ can
be singular and must be taken in the determining equations, not in
the solutions. This is true for any Hamiltonian in $E_2$ with an integral
of order $N$, independently of the separation of variables.
\item Two types of potentials occur which we call standard and
exotic. Standard ones are solutions of a linear compatibility
condition for the determining equations. For exotic potentials the linear
compatibility condition is satisfied trivially so the potentials satisfy nonlinear
equations. In quantum mechanics the nonlinear equations pass the Painlev\'e
test \cite{ARS, Conte}, in the classical case they do not.
\item The integrals of motion $H$, $X$ and $Y$ satisfy $[H, X] = [H, Y] = 0$, $[X, Y] = C \neq 0$,
where  $[\cdot, \cdot]$ denotes a Lie bracket in quantum mechanics
and a Poisson bracket in the classical case. Further commutations like
$[X,C]$, $[Y, C], \ldots,$ in general yield an infinite dimensional Lie algebra,
exceptionally a finite dimensional algebra \cite{Bargmann, Fock, Jauch},
or a Kac-Moody algebra \cite{DSD}. It
is however more fruitful to view this algebra as a finite dimensional polynomial
Lie or Poisson algebra \cite{Daskol, Milleretal, IanM}. In many cases
for $N=2, \ldots, 5$ it turns out that the commutators $[X,C] = D_1$, $[Y, C] = D_2$
are polynomials in $X$, $Y$ and $H$ with constant coefficients.
\item A simple observation is that if a potential $V(r, \theta)$ as in (\ref{Hpolar})
allows an integral $Y$ of order $N_0$ then the same potential will show up
again for infinitely many values of $N \geq N_0$. This is because all
powers $X^a Y^b + Y^b X^a$ and all commutators $[X^a, Y^b]$
(Lie or Poisson, respectively) are also integrals of motion. Only $3$ of them
(including the Hamiltonian) can be functionally (or polynomially) independent.
What is of interest is to determine the lowest order of $N$ for each
superintegrable potential. As an example we recall that Drach in his
pioneering article \cite{Drach1} found $10$
classical complex integrable potentials with a third-order integral (in $E_2( \mathbb {C})$).
Much later it was shown \cite{Ranada, Tsiganov:2000} that $7$ of the $10$ systems were
``reducible''. Indeed, these $7$ were actually second-order superintegrable and the
third-order integral is a commutator of two second-order ones.
\end{enumerate}


\section{$N^{th}$-order integral of motion: superintegrability}
\subsection{General $N^{th}$-order integral of motion}
In the {\bf classical case}, an additional general $N^{th}$-order polynomial integral has the form
\cite{Hietarinta1987, PostWinternitz:2015, Nikitin}
\begin{equation}\label{Y}
Y \ = \ Y^{(N)}  \ + \ \sum_{\ell=1}^{\big[\frac{N}{2}\big]}\sum_{j=0}^{N-2\ell} \,F_{j,2\ell}\ p_x^j\ p_y^{N-j-2\ell} \ ,
\end{equation}
where the functions $F_{j,2\ell}=F_{j,2\ell}(x,y)$ depend on the potential $V$ figuring in the Hamiltonian. In (\ref{Y}), the contributions of order $N$ are collected in the single term $Y^{(N)}$
\begin{equation}\label{YN}
Y^{(N)}  \ = \  \sum_{0\leq m+n\leq N}^{}\ A_{N-m-n,m,n}\ L_z^{N-m-n}\ p_x^m\,p_y^n  \ .
\end{equation}
Here $A_{N-m-n,m,n}$ are $\frac{(N+1)(N+2)}{2}$ constants. This leading term $Y^{(N)}$ (\ref{YN}) is fundamental since it defines the existence of an $N^{th}$-order integral. Hereafter we will focus on it.

Let us introduce the following notation in the $p$-plane $(p_x,\,p_y)$
\[
P \ \equiv \ \sqrt{p_x^2 \ + \ p_y^2} \quad ; \qquad
\tan \Phi \ \equiv \  \frac{p_y}{p_x} \ ,
\]
where the basis functions generating the irreducible representations of $SO(2)$ are given by
\[
{(p_x \, \pm \, i\, p_y)}^s\ =\ P^s\,({\cos s\,\Phi }\pm i\, {\sin s\,\Phi }) \ , \qquad  s= \ldots,-2,-1,0,1,2 \ldots \ .
\]

The $N^{th}$-order terms $Y^{(N)} $ (\ref{YN}) can be more conveniently written as
\begin{equation}\label{Ypolar}
Y^{(N)}  \ = \ \sum_{0\leq s+2k\leq N}^{}\,L_z^{N-s-2k}\,P^{s+2k}\,\bigg[ B^{(1)}_{N-s-2k,s,k}\,{\cos s\,\Phi } \ + \ B^{(2)}_{N-s-2k,s,k}\, {\sin s\,\Phi }    \bigg] \ ,
\end{equation}
where $B^{(\ell)}_{N-s-2k,\,s,\,k}$ ($\ell=1,2$) are $\frac{(N+1)(N+2)}{2}$ constants. Each of these constants
$B^{(\ell)}_{N-s-2k,\,s,\,k}$ can be expressed uniquely as a linear combination of the original parameters $A_{N-m-n,m,n}$ in (\ref{YN}).

For fixed $N,k$ and $s$, each pair
$
\big(B^{(1)}_{N-s-2k,s,k}\,\cos s\,\Phi ,\,B^{(2)}_{N-s-2k,s,k}\, {\sin s\,\Phi }\big)
$ for
($s \neq 0$) forms a doublet under $O(2)$ rotations. For $s=0$, the pair reduces to a singlet.

The integral $Y$ (\ref{Y}) is given in Cartesian coordinates for brevity and no separation of variables is assumed so far. To obtain the corresponding expression in polar coordinates we put
$p_x \ = \ \cos \theta \,p_r - \frac{\sin \theta}{r}\,L_z$, $p_y \ = \ \sin \theta \,p_r + \frac{\cos \theta}{r}\,L_z$.

Similarly, in the {\bf quantum case} \cite{PostWinternitz:2015, Nikitin}
we can write the Hermitian $N^{th}$-order operator $Y$
\begin{equation}\label{Yquatum}
Y \ = \ Y^{(N)}  \ + \ \sum_{\ell=1}^{\big[\frac{N}{2}\big]}\sum_{j=0}^{N-2\ell} \,\{F_{j,2\ell},\,\ p_x^j\ p_y^{N-j-2\ell}\} \ ,
\end{equation}
\begin{equation}\label{Yquantumpolar}
Y^{(N)}  \ = \ \sum_{0\leq s+2k\leq N}^{}\,\bigg\{L_z^{N-s-2k}, \ (p_x^2 + p_y^2)^{k}\,\bigg( B^{(1)}_{N-s-2k,s,k}\,\bigg[{(p_x + i\,p_y) }^s\bigg]_{Re} \ + \ B^{(2)}_{N-s-2k,s,k}\,\bigg[{(p_x + i\,p_y) }^s\bigg]_{Im}   \bigg) \bigg\} \ .
\end{equation}
Above $F_{j,2\ell}=F_{j,2\ell}(x,y)$ depends on the potential $V$ and $[c\,]_{Re}$ refers to the real part of $c$ where formally we treat the operators $p_x$ and $p_y$ as variables. The
$B^{(\ell)}_{N-s-2k,\,s,\,k}$ ($\ell=1,2$) are constants, and $\{\,,\,\}$ denotes an anticommutator.

\subsection{Determining equations}
In the \textbf{quantum case}, since $Y$ is an $N^{th}$-order operator, the commutator $[H,\,Y]$ is an
operator of order $(N+1)$, {\it i.e.} we have
\begin{equation}\label{HYC}
  [H,\,Y] \ = \ \sum^{N+1}_{k+l=0} Z_{k,l}(r,\,\theta)\,\frac{\partial^{k+l}}{\partial r^k\,\partial \theta^l} \ .
\end{equation}
We require $Z_{k,l}=0$ for all $k$ and $l$ and obtain the
determining equations. The terms of order $k+l=N+1$ and $k+l=N$ vanish automatically.
This was already shown\cite{PostWinternitz:2015} for arbitrary $N$. Moreover, only the terms with $k+l$ having
the opposite parity than $N$ provide independent determining equations ($Z_{k,l}=0$). Those with the same parity provide equations that are differential consequences of the first ones \cite{PostWinternitz:2015}. For the \textbf{classical case}, the determining equations can be obtained from the quantum case by taking the limit $\hbar\rightarrow 0$\,.

\subsection{Linear Compatibility Condition}
It has been shown \cite{Hietarinta1987, PostWinternitz:2015} that vanishing of the Poisson or Lie bracket
$[H, Y] = 0$ implies that the potential $V$ must satisfy a linear PDE of order $N$, a linear compatibility condition (LCC)
for the determining equations $Z_{k,l}=0$ for $k+l=N-1$ . For arbitrary potentials this linear PDE takes the form
\begin{equation}\label{LCC}
\sum_{j=0}^{N-1}{(-1)}^j\,\partial_x^{N-1-j}\,\partial_y^{j}\,\big[ (j+1)\,f_{j+1,0}\,\partial_x V\ + \ (N-j)\,f_{j,0}\,\partial_yV  \,     \big]  \ = \ 0   \ .
\end{equation}
This LCC (\ref{LCC}) is a necessary (but not sufficient) condition for the existence of the integral $Y$. The functions $f_{j,0}$ do not depend on the potential
\begin{equation}\label{fj0}
f_{j,0} \ = \  \sum_{n=0}^{N-j}\sum_{m=0}^{j}\binom{N-m-n}{j-m}\ A_{N-m-n,m,n}\ x^{N-j-n}\ {(-y)}^{j-m}   \ ,
\end{equation}
and they are completely determined by the coefficients $A_{N-m-n,m,n}$ of $Y^{(N)}$ (\ref{YN}). The LCC does not contain $\hbar$  and thus it is the same for both classical and quantum systems. The LCC will play a fundamental role in the present description of the superintegrable systems.

\subsection{Standard and Exotic potentials}
In this section we impose superintegrablity, namely the condition $[H,\, Y] = 0$. This will determine
the functions $R(r)$ and $S(\theta)$. The most general equation for the angular part $S(\theta)$ corresponds
to the case $R(r)=0$.

In the \textbf{classical case}, let us split the $N^{th}$-order terms $Y^{(N)}$ (\ref{Ypolar}) in the integral $Y$ into two parts
\begin{equation}\label{YNg}
Y^{(N)} \ = \ Y_{I}^{(N)} \ + \ Y_{II}^{(N)} \ ,
\end{equation}
\begin{equation}\label{YI}
Y_I^{(N)} \ = \ \sum_{  N-1\leq s+2k\leq N}^{}\,L_z^{N-s-2k}\,P^{s+2k}\,\bigg[ B^{(1)}_{N-s-2k,s,k}\,{\cos s\,\Phi } \ + \ B^{(2)}_{N-s-2k,s,k}\, {\sin s\,\Phi }    \bigg]  \ ,
\end{equation}
\begin{equation}\label{YII}
Y_{II}^{(N)} \ = \ \sum_{0\leq s+2k\leq N-2}^{}\,L_z^{N-s-2k}\,P^{s+2k}\,\bigg[ B^{(1)}_{N-s-2k,s,k}\,{\cos s\,\Phi } \ + \ B^{(2)}_{N-s-2k,s,k}\, {\sin s\,\Phi } \bigg]    \ .
\end{equation}

In general, the Hamiltonian $H$ is not invariant under dilations $(x,y)\rightarrow (\sigma\,x,\,\sigma\,y)$. However, for $R(r)=0$
it does scale as $H \rightarrow \frac{1}{\sigma^2}H$. Also, under dilations
\[
Y_I^{(N)} \ \rightarrow \  \sum_{  N-1\leq s+2k\leq N}^{}\frac{1}{\sigma^{s+2k}}\,L_z^{N-s-2k}\,P^{s+2k}\,\bigg[ B^{(1)}_{N-s-2k,s,k}\,{\cos s\,\Phi } \ + \ B^{(2)}_{N-s-2k,s,k}\, {\sin s\,\Phi }    \bigg] \ ,
\]
\[
Y_{II}^{(N)} \ \rightarrow \  \sum_{0\leq s+2k\leq N-2}^{}\frac{1}{\sigma^{s+2k}}\,L_z^{N-s-2k}\,P^{s+2k}\,\bigg[ B^{(1)}_{N-s-2k,s,k}\,{\cos s\,\Phi } \ + \ B^{(2)}_{N-s-2k,s,k}\, {\sin s\,\Phi } \bigg] \ .
\]

The LCC (\ref{LCC}) was obtained \cite{PostWinternitz:2015} only from the determining equations corresponding to the highest order terms of the commutator $[H,Y]$, namely those that scale as $\frac{1}{\sigma^{N-1}}$ and $\frac{1}{\sigma^{N-2}}$. For $R(r)=0$, the system becomes scale-invariant and then the LCC equation must be independent of the lower order terms $\frac{1}{\sigma^{\alpha}}$, $\alpha=N-3,N-4,\ldots,0$, otherwise the equation $[H,Y]=0$ would not be invariant under dilations. Therefore, the constants $B^{(1)}_{N-s-2k,s,k}$ and $B^{(2)}_{N-s-2k,s,k}$ figuring in $Y_{II}^{(N)}$ do not appear in the LCC (\ref{LCC}) while those in $Y_{I}^{(N)}$  do occur in the LCC. For $R(r)=0$, the compatibility condition is satisfied identically for $Y_{I}^{(N)}=0$. Moreover, the terms $Y_I^{(N)}$ and $Y_{II}^{(N)}$ divide into subclasses that are characterized by the exponent $\beta$ of the scaling $\frac{1}{\sigma^{\beta}}$.

For the \textbf{quantum case}, the analogs of (\ref{YI}) and (\ref{YII}) read
{\small
\begin{equation}\label{YINnew}
Y_I^{(N)} \ = \ \sum_{N-1\leq s+2k\leq N }^{}\,\bigg\{L_z^{N-s-2k},\ P^{2k}\,\bigg( B^{(1)}_{N-s-2k,s,k}\,\bigg[{(p_x + i\,p_y) }^s\bigg]_{Re} \ + \ B^{(2)}_{N-s-2k,s,k}\,\bigg[{(p_x + i\,p_y) }^s\bigg]_{Im}   \bigg)\bigg\}  \ ,
\end{equation}
\begin{equation}\label{YIINnew}
Y_{II}^{(N)} \ = \ \sum_{ 0\leq s+2k\leq N-2;}^{}\,\bigg\{L_z^{N-s-2k},\ P^{2k}\,\bigg( B^{(1)}_{N-s-2k,s,k}\,\bigg[{(p_x + i\,p_y) }^s\bigg]_{Re} \ + \ B^{(2)}_{N-s-2k,s,k}\, \bigg[{(p_x + i\,p_y) }^s\bigg]_{Im} \bigg)\bigg\}    \ .
\end{equation}
}

Thus, for both classical and quantum systems we distinguish two cases for $R(r)=0$:
\begin{itemize}
  \item I) $Y_{I}^{(N)}\neq 0$: this case corresponds to \emph{Standard potentials} for which the angular component $S(\theta)$ satisfies the LCC (\ref{LCC}).
  \item II) $Y_{I}^{(N)} = 0$: this situation corresponds to the \emph{Exotic potentials}. All coefficients
  in (\ref{LCC}) satisfy $f_{a,0} = 0$ and
the function $S(\theta)$ is not constrained by this linear equation.
\end{itemize}

The reflection operator of the radial variable ($r \rightarrow -r$, or equivalently $\theta \rightarrow \theta + \pi $) commutes with the Hamiltonian $H$ as well. Therefore, we can use both symmetries namely the parity transformation and the scaling to classify the $N^{th}$-order terms $Y^{(N)}$ (\ref{Ypolar}) of the integral $Y$. Each class will be associated
with different types of potentials.

\subsection{Trivial integrals}
$N^{th}$-order integrals exit both in the classical and quantum systems
and they are related to $O(2)$ singlets. In the classical case, they occur for $s=0$ in (\ref{YNg})
and are given by
$
Y^{(N)}_{singlets} \ = \  \sum_{k = 0}^{}B^{(1)}_{N-2k,0,k} \,L_z^{N-2k}\, P^{2k} \,.
$
For even $N$, by taking linear combinations of the form ($Y
+ \sum_{0 \leq i+j\leq N}^{} a_{ij}\, X^{i}\,H^{j}$), where the $a_{ij}$ are constants, we
can eliminate $Y_{singlets}^{(N)}$, we are just adding or subtracting trivial integrals made out of $X$ and $H$.
For odd $N$ ($N = 2 k +1$) the situation is more complicated. For $N=3$ and $N=5$ one of two possibilities can
occur. The first is that the singlets lead to functions $S(\theta)$ expressed in terms of Weierstrass elliptic functions.
In these cases a syzygy (a polynomial relation) $P(X, Y, H) = 0$ exists between the $3$ integrals so the system
is not really superintegrable. The second possibility is that we reobtain already known lower order superintegrable systems
with $Y$ expressed in terms of $X, H$ and $Y^{N^{\prime}}$ with $N^{\prime} < N$. Our calculations
indicate that these are the only options for all values of $N = 2 k + 1$, $k \in \mathbb{N}$. For $k = 0$ this
statement corresponds to Burchnall-Chaundy theory \cite{Bch}.

For the classical case, any $N^{th}$-order integral ${Y_0}$ for which the leading term ${Y_0^{(N)}}$ is polynomially related with the trivial/lower order integrals
\begin{eqnarray}
{\big(Y_0^{(N)}\big)}^l \ = \  X^m\,H^n\,{\big(Y^{(N-j)}\big)}^k \ + \ \text{lower order terms} \ ,
\quad l,m,n,k,j < N \in \mathbb{Z}_+\,,
\label{Y0}
\end{eqnarray}
will lead to an already known ({\it i.e.}; lower order)
superintegrable system.

Notice that for $k=0$, such an integral ${Y_0}$ is a trivial one since $H$ and $X$ are conserved independently of the potential $V$ (\ref{Hpolar}). In the case $k\neq 0$, ${Y_0}$ is an integral for specific potentials only. The extension of (\ref{Y0}) to the quantum case is straightforward. We mention that such ``reducible'' integrals for $k>0$ may be useful for calculating trajectories, or
solving spectral problems.

\section{Radial component $R(r)$}
Our entire approach to finite-dimensional integrable and superintegrable
systems is based on the assumption that a fundamental set of integrals
of motion exists that are all polynomials of some finite degree $N$ in the
momenta. In this article we are considering the case of a two-dimensional
Euclidean space $E_2$ so at most $3$ independent integrals can exist:
$H$, $X$ and $Y$ of (\ref{Hpolar}), (\ref{X}) and (\ref{Y}), respectively. We
have presented the determining equations for the coefficients $Z_{k,l}$ in (\ref{HYC})
and the linear compatibility condition (\ref{LCC}) for a subset of these equations.

Eq. (\ref{LCC}) can be written in polar coordinates for the separable potential (\ref{Hpolar}) as
{\footnotesize
\begin{equation}
\begin{aligned}
& \sum_{j=0}^{N-1}{(-1)}^j\,{\big(\cos \theta \,\partial_r - \frac{\sin
\theta}{r}\,\partial_\theta \big)}^{N-1-j}\,{\big( \sin \theta
\,\partial_r + \frac{\cos \theta}{r}\,\partial_\theta  \big)}^{j}\,  \bigg[
[\,(j+1)\,f_{j+1,0}\,\cos \theta +  (N-j)\,f_{j,0}\,\sin \theta\,]\,R'
\\ &  -\frac{2}{r^3}[\,(j+1)\,f_{j+1,0}\,\cos \theta +  (N-j)\,f_{j,0}\,\sin \theta\,]\,S + \frac{1}{r^3}[\,(N-j)\,f_{j,0}\,\cos \theta -(j+1)\,f_{j+1,0}\,\sin \theta ]\,S' \bigg] \ = \ 0   \ .
\label{LCCinpolar}
\end{aligned}
\end{equation}
}
Multiplying (\ref{LCCinpolar}) by $r^{(N+2)}$ and differentiating it twice with respect to $r$ we obtain
an equation involving $R(r)$ with a polynomial dependence on $r$ and a trigonometric dependence on
$\theta$. The function $S(\theta)$ has been eliminated by the derivatives with respect to $r$. Separating terms proportional to $\sin s \theta$ and $\cos s \theta$ $(s = 1, \dots, N)$ we get a system of $2 N$ linear ODEs of order $(N+2)$ for the radial function
$R(r)$ alone. These $2N$ equations vanish trivially only if all coefficients in the leading part $Y^{(N)}$
of the integral $Y$ themselves vanish, {\it i.e.} $Y$ is of lower order. If $R(r)$ satisfies $R(r) \equiv 0$ then
the LCC (\ref{LCCinpolar}) greatly simplifies and can be used to determine $S(\theta)$. This corresponds
to standard potentials (standard potentials also appear for $R(r) \neq 0$).

We have also checked that for $N=2,3,4,5$, the $2N$ equations are solved nontrivially ({\it i.e.} $Y^{(N)} \neq 0$)
for two confining potentials with
\begin{equation}\label{Rp}
R(r) \ = \ \frac{a}{r}  \quad {\rm and} \quad  R(r) \ = \ b\,r^2     \ ,
\end{equation}
though these are not necessarily the only solutions. However, only these two (\ref{Rp}) are compatible with all the determining equation $Z_{k,l}=0$ in (\ref{HYC}).

A further remark is in order, the relation between maximal superintegrability ($(2\,n-1)$ independent well-defined integrals of motion for $n$ degrees of freedom) and stable closed orbits in classical mechanics was established by Nekhoroshev\cite{Nekhoroshev}. In classical mechanics the property that all bounded trajectories are closed is an equivalent definition of a maximally superintegrable system (at least for
confining potentials).

In a remarkable article, Onofri and Pauri\cite{Onofri} generalized the Bertrand theorem \cite{Bertrand1873, Goldstein} and
established that for $S(\theta) \neq 0$ in the separable potential (\ref{Hpolar}) all bounded trajectories are closed (for
certain functions $S(\theta)$) for $R(r) = b r^2$ and
\begin{equation}\label{Rgo}
   R(r) \ =\ \frac{1}{r^2}\sqrt{a^2\,r^2\,+\,d} \ .
\end{equation}
Thus, the Kepler-Coulomb potential is recovered from a family of superintegrable potentials
for $d = 0$. This result was recently confirmed and generalized to all two-dimensional spaces
of constant curvature by Gonera and Kaszubska\cite{Gonera2}.

The difference between their approach and ours is that we request the additional
integral $Y$ to be a polynomial of (arbitrary) order $N$ in the momenta. They only
require that a globally defined integral should exist. Their function $R(r)$ is more general
but the arguments only hold in classical mechanics. Our results for $R(r)$ as in (\ref{Rp})
hold in both classical and quantum mechanics. Moreover, in quantum mechanics they lead
to the existence of potentials expressed in terms of Painlev\'e transcendents.

It can be shown that the function $R(r)$ of (\ref{Rgo}) with $d \neq 0$ does not
satisfy any of the $2 N$ linear ODEs for the radial function
$R(r)$ obtained from (\ref{LCCinpolar}). Details and the general case
will be presented elsewhere.

\section{Angular component $S(\theta)$}

\subsection{Classical Systems}

\textbf{Standard potentials}. For $Y_{I}^{(N)}\neq 0$, the potential (\ref{Hpolar})
must satisfy the LCC (\ref{LCCinpolar}). For $R(r)$ restricted to one of the cases (\ref{Rp}),
(\ref{LCCinpolar}) reduces to a linear ODE for $S(\theta)$.
\begin{itemize}
\item For even $N$, the standard potentials, {\it i.e.}; the non-trivial solutions of the LCC (\ref{LCCinpolar}), are the TTW potential \cite{TTWclassical} and the PW potential \cite{PostWinternitz:2010}
\begin{eqnarray}
V_{\rm TTW}\ =\ b\, r^2\, +\, \frac{1}{r^2} \left[\frac{\alpha}{\cos^2 (k \theta)}\, +\,
\frac{\beta}{\sin^2 (k \theta)}\right]\,,
\,
V_{\rm PW} \ = \ \frac{a}{r} \ + \ \frac{1}{r^2}\left[\frac{\mu}{\cos^2(\frac{k}{2} \theta)}\, +\, \frac{\nu}{\sin^2(\frac{k}{2} \theta)} \right]\,,
\nonumber
\end{eqnarray}
where $k=m/n$ and $m$ and $n$ are two integers (with no common divisors). In the TTW case, $N \,= \,2 \,(m + n - 1)$.
The case $R(r) = 0$ is reobtained for $a=0$ or $b=0$.

\item For odd $N$, the radial part vanishes $R(r) = 0$.
\end{itemize}

\textbf{Exotic potentials}. These potentials correspond to the case $Y_{I}^{(N)}=0$, $Y_{II}^{(N)}\neq0$. For $N=3,4$, examples of classical exotic potentials were presented earlier \cite{TPW2010, AMJVPW2015}. Up to a simple change of variables the angular part $S(\theta)$ satisfies exactly the same non-linear first-order differential equation in both cases $N=3,4$. We have checked that for $N=5$ the potential also obeys this equation. In particular, for $N=3,4,5$  there exist superintegrable potentials $V=b\,r^2 + \frac{T'(\theta)}{r^2}$ where
\[
T(z) \ \propto \ \frac{z^{\frac{1}{3}}{(3\,z^2+2\sqrt{4+3\,z^2}+5)}^{\frac{1}{6}}}{{(\sqrt{4+3\,z^2}+2)}^{\frac{2}{3}}} \ ,
\qquad
z=\tan [(N-2)\theta]\,.
\]
The results for $N=3, 4, 5$, lead us to the following conjecture:
\begin{conjecture}
For $Y_{I}^{(N)}=0$, $Y_{II}^{(N)}\neq0$, superintegrable classical systems appear such that
\begin{itemize}
\item  For even $N>2$, the potentials are given by the deformed Kepler potential
or the deformed harmonic oscillator
\begin{equation}\label{Vaabb}
V \ = \ \frac{a}{r}  \ + \ \frac{T'(\theta)}{r^2}    \ ,
\qquad
V \ = \ b\,r^2  \ + \ \frac{\tilde T'(\theta)}{r^2} \ ,
\qquad
S(\theta) \ \equiv \ T'(\theta) \ \, \text{or} \ \, \tilde T'(\theta) \ ,
\end{equation}
where the angular components $T(\theta)$ and $\tilde T(\theta)$ obey a non-linear ODE that does not pass the Painlev\'e test. The functions $\tilde{T}$ and $T$ depend parametrically on the integer $N$. For odd $N$ the radial part vanishes.
\item In particular, there exists an infinite family of superintegrable potentials $V=b\,r^2 + \frac{T'(\theta)}{r^2}$, labeled by $N$, having the leading term of the corresponding integral $Y$ as $\{L_z^2,\,\big[{(p_x + i\,p_y) }^{N-2}\big]_{Re} \}$. In this case the function $T(\theta)$ obeys the following non-linear first order ODE
\[
3\,\tau^2\,(\tau^2+1){(T')}^2 \ + \ 4\,\tau\,(c_1\,\tau+c_2)\,T' \ + \ 2\,\tau\,T\,T' \ - \ T(T+4\,c_2) \ + \ \frac{c_3}{\sqrt{\tau^2+1}} \ + \ c_4 \ = \ 0 \ ,
\]
where
\begin{equation}
\tau\ \equiv \ \bigg\{\begin{array}{ll}
    \cos^2[\frac{1}{2}(N-2)\,\theta ]\\
    \sin^2[\frac{1}{2}(N-2)\,\theta ]\ , \\
    \end{array}
\end{equation}
and $c_{1,2,3,4}$ are arbitrary constants. For odd N the constants satisfy $b=c_3=0$.
\end{itemize}
\end{conjecture}

\subsection{Quantum systems}

\textbf{Standard potentials}. For $Y_{I}^{(N)}\neq 0$, $N^{th}$-order quantum superintegrable systems do occur such that
\begin{itemize}
\item For even $N$, the standard potentials, {\it i.e.}; the solutions of the LCC (\ref{LCCinpolar}), in addition to $R(r)=0$ allow two confining potentials. These are the deformed Kepler potential and the deformed harmonic oscillator
\begin{equation}
V \ = \ \frac{a}{r}  \ + \ \frac{ T'(\theta)}{r^2}  \  ,
\qquad
V \ = \ b\,r^2  \ + \ \frac{\tilde T'(\theta)}{r^2}  \ .
\end{equation}
In the first case we have
\begin{equation}\label{Vacl}
{T}(\theta) \ = \frac{\alpha_0 \ + \  \sum_{s=1,3,5,...}^{N-1}  ( \alpha_s\,\cos s\, \theta \, + \, \beta_s\,\sin s\, \theta )}
{\sum \limits_{s=1,3,5,...\,;\,N-1\leq s+2k\leq N}^{}  \big( B^{(1)}_{N-s-2k,\,s,\,k}\, {\cos s\,\theta }\, + \, B^{(2)}_{N-s-2k,\,s,\,k}\,{\sin s\,\theta }\big)}\ ,
\end{equation}
where the $\alpha$'s and the $\beta$'s are constants which in general depend on the Planck's constant $\hbar^2$ and the $B^{(\ell)}_{N-s-2k,\,s,\,k}$ ($\ell=1,2$) in such a way that new pure quantum potentials \cite{Hietarinta1998} proportional to $\hbar^2$ occur with no classical counterpart. Such constants are fixed by requiring the potential to satisfy the determining equations (see (\ref{HYC})).

In the second case we have
\begin{equation}\label{Vbcl}
\tilde T(\theta) \ = \frac{\tilde \alpha_0 \ + \  \sum_{s=2,4,...}^{N}  (  \tilde \alpha_s\,\cos s\, \theta \,  + \, \tilde \beta_s\,\sin s\, \theta  )}
{\sum \limits_{s=2,4,...\,;\, N-1\leq s+2k\leq N}^{} s\,\big(B^{(2)}_{N-s-2k,\,s,\,k}\,{\cos s\,\theta } \ - \ B^{(1)}_{N-s-2k,\,s,\,k}\, {\sin s\,\theta} \big)} \ .
\end{equation}

\item For odd $N \geq 3$ the radial component is zero $R(r)=0$. In this case the angular component is given by
\begin{equation}\label{Vaclodd}
{T}(\theta) \ =  \frac{\sum_{s=1,3,5,...}^{N}  ( \alpha_s\,\cos s\, \theta \, + \, \beta_s\,\sin s\, \theta )}
{\sum \limits_{s=1,3,5,...\,;\,N-1\leq s+2k\leq N}^{} s\, \big( B^{(2)}_{N-s-2k,\,s,\,k}\, {\cos s\,\theta }\, - \, B^{(1)}_{N-s-2k,\,s,\,k}\,{\sin s\,\theta }\big)} \ ,
\end{equation}
\begin{equation}\label{Vbclodd}
\tilde T(\theta) \ = \frac{\tilde \alpha_0 \ + \  \sum_{s=2,4,...}^{N - 1}  (  \tilde \alpha_s\,\cos s\, \theta \,  + \, \tilde \beta_s\,\sin s\, \theta  )}
{\sum \limits_{s=2,4,...\,;\, N-1\leq s+2k\leq N}^{} \big(B^{(1)}_{N-s-2k,\,s,\,k}\,{\cos s\,\theta } \ + \ B^{(2)}_{N-s-2k,\,s,\,k}\, {\sin s\,\theta} \big)} + \tilde{\alpha}_{N+1} \ .
\end{equation}

\item Both the quantum TTW and PW systems are fully contained
in the family $Y_{I}^{(N)}\neq 0$, as particular cases.

\end{itemize}

\bigskip

\textbf{Exotic potentials}. These potentials correspond to the case $Y_{I}^{(N)}=0$, $Y_{II}^{(N)}\neq0$. For $N=3,4$, examples of quantum exotic potentials were also presented in \cite{TPW2010} and \cite{AMJVPW2015}, respectively. Upon a simple change of variables the angular part $S(\theta)$ satisfies exactly the same non-linear second-order differential equation in both cases $N=3,4$. Its general solution can be written in terms of the sixth Painlev\'e transcendent $P_6$. The transcendent $P_6$ occurs for the case $N=5$ as well.

\begin{conjecture}
For $Y_{I}^{(N)}=0$, $Y_{II}^{(N)}\neq0 $, quantum superintegrable systems occur where
\begin{itemize}
\item  For even $N$, the radial part in addition to $R(r) = 0$ can be either $R(r) = \frac{a}{r} $ or $R(r)=b\,r^2$. In this case the general form of $T(\theta)$ is given by the solution of a non-linear ODE which passes the Painlev\'e test. For odd $N$, the radial part is only $R(r)=0$.
\item  In particular, there exists an infinite family of superintegrable potentials $V=b\,r^2 + \frac{T'(\theta)}{r^2}$, labeled by $N$, with the leading term of the corresponding integral $Y$ as $\{L_z^2,\,\big[{(p_x + i\,p_y) }^{N-2}\big]_{Re} \}$. The function $T(\theta)$ is given by
\begin{equation}\label{SP6}
T(\tau)\ = \ \hbar^2\,(N-2)\, \bigg[\frac{W(\tau)}
{\sqrt{\tau}\sqrt{1-\tau}}
\ + \ \gamma\,\frac{ (1-2\,\tau)}
{ 4\, \sqrt{\tau}\,\sqrt{1-\tau}} \bigg] \ ,
\end{equation}
where
\begin{equation}
\tau\ \equiv \ \bigg\{\begin{array}{ll}
    \cos^2[\frac{1}{2}(N-2)\,\theta ]\\
    \sin^2[\frac{1}{2}(N-2)\,\theta ]\ , \\
    \end{array}
\end{equation}

\begin{align}
 W(\tau\,;\,\gamma_1,\,\gamma_2,\,\gamma_3,\,\gamma_4) \ & =\ \frac{\tau^2(\tau-1)^2}{4P_6(P_6-1)(P_6-\tau)}\bigg[P_6'-\frac{P_6(P_6-1)}{\tau(\tau-1)}
\bigg]^2+\frac{1}{8}(1-\sqrt{2\gamma_1})^2(1-2P_6)\nonumber\\
&-\frac{1}{4}\gamma_2\bigg(1-\frac{2\tau}{P_6}\bigg)-\frac{1}{4}
\gamma_3\bigg(1-\frac{2(\tau-1)}{P_6-1}\bigg)+\bigg(\frac{1}{8}-\frac{\gamma_4}{4}
\bigg)\bigg(1-\frac{2z(P_6-1)}{P_6-\tau}\bigg) \ ,
\label{Wpot}
\end{align}
with $\gamma_1,\gamma_2,\gamma_3$ and $\gamma_4$ the parameters that define the sixth Painlev\'e
transcendent $P_6$ which satisfies the well known second-order differential
equation:
\begin{eqnarray}
\label{P6}
 P_6''=\frac{1}{2}\bigg[\frac{1}{P_6}+\frac{1}{P_6-1}+\frac{1}{P_6-\tau}\bigg]
(P_6')^2-\bigg[\frac{1}{\tau}+\frac{1}{\tau-1}+\frac{1}{P_6-\tau}\bigg]P_6'\nonumber\\
+\frac{P_6(P_6-1)(P_6-\tau)}{\tau^2(\tau-1)^2}\bigg[\gamma_1+\frac{\gamma_2\,\tau}{P_6^2}
+\frac{\gamma_3\,(\tau-1)}{(P_6-1)^2}+\frac{\gamma_4\,\tau(\tau-1)}{(P_6-\tau)^2}\bigg] \ ,
\end{eqnarray}
$\gamma=(\gamma_2+\gamma_4)-(\gamma_1+\gamma_3)+\sqrt{2\,\gamma_1}-\frac{3}{4}$. For $N$ even, all four $\gamma_1,\gamma_2,\gamma_3$ and $\gamma_4$ are arbitrary. For $N$ odd, $b=0$ and the $\gamma$'s are related as $(\gamma_2+\gamma_3)(\gamma_1+\gamma_4-\sqrt{2\gamma_1})=0$ .
\end{itemize}
\end{conjecture}

\section{CONCLUSIONS}
\label{Conclusions}
We considered superintegrable systems in a two-dimensional Euclidean plane. Classical
and quantum $N^{th}$-order superintegrable potentials separating in polar
coordinates were classified into two categories. This classification is based on the LCC (\ref{LCCinpolar}), a linear compatibility condition for the existence of the $N^{th}$-order integral $Y$. Correspondingly, the leading terms of $Y$ split into two parts $Y_I^{(N)}$ and $Y_{II}^{(N)}$. The first part $Y_I^{(N)}$ contains at most linear terms in the angular momentum $L_z$ whereas in $Y_{II}^{(N)}$ only higher order terms $L_z^{m}$, $m=2,3,\ldots,N$, occur.
Firstly, unlike the angular part $S(\theta)$ the radial function $R(r)$ always satisfies a linear differential equation. The Kepler and Harmonic radial parts ({\it i.e.}; $\frac{a}{r}$ and $b\,r^2$, respectively) are the only ones that admit an $N^{th}$-order polynomial integral. Secondly, the most general form of the angular part $S(\theta)$ corresponds to the non-confining case $R(r)=0\,$. For this case, the Hamiltonian possesses additional symmetries, it becomes scale and reflection invariant and the LCC (\ref{LCCinpolar}) becomes independent of $Y_{II}^{(N)}$. The above-mentioned emergent symmetries lead to a further classification of the potentials into different subclasses. We can summarize the main results via the following Theorems and Conjectures

{\bf Theorem 1.} \emph{For the standard potentials $Y_{II}^{(N)}=0$, the angular part $S(\theta)$ satisfies a linear differential equation (\ref{LCCinpolar}). There are two mutually exclusive forms of such standard potentials, corresponding to (\ref{Vacl}) and (\ref{Vbcl}), respectively. The leading term of the integral $Y$ is given
by (\ref{YI}) and (\ref{YINnew}) for the classical and quantum systems, respectively. The classical standard potentials coincide with the} $V_{\rm TTW}$ \emph{and} $V_{\rm PW}$ \emph{potentials}. \emph{In the quantum case new pure quantum potentials appear in addition to the} $V_{\rm TTW}$ \emph{and} $V_{\rm PW}$.

{\bf Theorem 2.} \emph{For the exotic potentials $Y_{I}^{(N)}=0$, the linear differential equation (\ref{LCCinpolar}) vanishes identically. The leading term of the integral $Y$ is given by (\ref{YII}) and (\ref{YIINnew}) for the classical and quantum systems, respectively.}

Complementary, based on the results for $N=3,4,5$, we conjecture (\textbf{Conjecture 1}) that for classical exotic potentials the angular part $S(\theta)$ satisfies a non-linear ODE which does not pass the Painlev\'e test. An infinite family of such potentials was presented.

For the quantum exotic potentials we conjecture (\textbf{Conjecture 2}) that $S(\theta)$ satisfies a non-linear ODE which passes the Painlev\'e test. Here, a new infinite family of exotic quantum superintegrable potentials in terms of the sixth Painlev\'e transcendent $P_6$ was introduced as well.

Work is currently in progress on a continuation of this article. We will
present an extended paper with more explicit examples of the classification just mentioned above.
For the cases $N=3,4,5$ we plan to present the polynomial algebra generated by the
integrals of motion and to use it to calculate the energy spectrum and the wave
functions in the quantum case. The search of an underlying hidden algebraic structure of the Hamiltonian is also relevant towards its exact solvability. A similar classification for potentials separating in Cartesian
coordinates will be presented elsewhere.

\section{ACKNOWLEDGMENTS}
We thank an anonymous referee for calling the article\cite{Onofri} by Onofri and Pauri to our attention.
The research of AME was partially supported by a fellowship awarded
by the Laboratory of Mathematical Physics of the CRM. The research
of PW was partially supported by a research grant from NSERC of Canada.
The research of \.{I}Y was partly supported by Hacettepe University Scientific Research
Coordination Unit. Project Number: FBI-2017-14035. He also thanks the Centre de
Recherches Math\'{e}matiques, Universit\'{e} de Montr\'{e}al for kind hospitality during his sabbatical leave.

{}

\end{document}